\documentclass[useAMS,usenatbib]{mn2e}
\usepackage{graphicx}
\usepackage{amssymb}
\title[Detecting and characterizing interstellar bubbles with FPS]
{BUBBLY\thanks{The code used in this paper can be found at https://github.com/artemic/Bubbly.git}: 
A method for detecting and characterizing interstellar bubbles using Fabry-Perot spectroscopy}

\author[A. Camps-Fari\~{n}a et al.]{A. Camps-Fari\~{n}a$^{1,2}$\thanks{E-mail:artemic@iac.es}, 
J. Zaragoza-Cardiel$^{1,2}$, J.E. Beckman$^{1,2,3}$, J. Font$^{1,2,4}$,\newauthor 
B. Garc\'{i}a-Lorenzo$^{1,2}$, S. Erroz-Ferrer$^{1,2}$, 
P. Amram$^{5}$\\$^{1}$Instituto de Astrof\'{i}sica de Canarias, C/V\'{i}a L\'{a}ctea s/n, E-38205 La Laguna, 
Tenerife, Spain\\
$^{2}$Department of Astrophysics, University of La Laguna, E-38205 La Laguna, Tenerife, Spain\\
$^{3}$CSIC, 2806 Madrid, Spain\\
$^{4}$National Observatory of Athens\\
$^{5}$Laboratoire d'Astrophysique de Marseille, Aix Marseille Universit\'{e}, CNRS, F-13388, Marseille, 
France}

\begin{document}

%\date{Accepted 1988 December 15. Received 1988 December 14; in original form 1988 October 11}

\pagerange{\pageref{firstpage}--\pageref{lastpage}} \pubyear{2002}

\maketitle

\label{firstpage}

\begin{abstract}
We present a new method for the detection and characterization of expansion in galaxy discs based on 
H$\alpha$ Fabry-Perot spectroscopy, taking advantage of the high spatial and velocity resolution of 
our instrument (GH$\mathrm{\alpha{}}$FaS). The method analyses multi-peaked emission line profiles to 
find expansion along the line of sight on a point-by-point basis.
At this stage we have centred our attention on the large scale structures of expanding gas associated 
with HII regions which show a characteristic pattern of expansion velocities, of order 100 km/s, as a 
result of both bubble shape and projection effects. We show an example of the expansion map obtained 
with our method from a superbubble detected in the Antennae galaxies. We use the information obtained 
from the method to measure the relevant physical parameters of the superbubbles, including their ages 
which can be used to date young star clusters. 
\end{abstract}


\section{Introduction}
The study of multi-component emission lines is a topic of major interest as they are produced by many systems in astrophysics. It has long been known that the spectrum of an AGN has two components to its emission, characterised by a narrow and a broad peak each associated with two parts of the system each having very different kinematics \citep{antonucci93}. The presence of an underlying broad supersonic component, which we now know corresponds to unresolved kinematic components, is also characteristic of bright HII regions \citep{arsenault86,chu94,yang96}. In this case the  observed emission peak may in fact be a combination of the main peak and two unresolved peaks associated with an expanding shell. Multiple peaks also appear in CO observations when independent clouds overlap in the line of sight \citep{blitz86,stutzki90}, and outflows of gas have been observed in ionized and molecular gas \citep{heckman90,lada85,bally83,bland88}. It is apparent that the study of these multiple components of an emission line, related to different kinematic phenomena, is relevant to a wide range of astrophysical problems.

Massive stars in OB associations and star-forming regions inject large quantities of mechanical energy into the surrounding interstellar medium in the form of intense winds and supernova explosions that sweep up the material around them. This creates an expanding shell of dense material leaving a cavity of hot low density gas inside. The individual bubbles created by each star combine quickly into one big superbubble associated with the region. A definition by \cite{chu08} by which a bubble is blown by the stellar winds while a superbubble also has input from supernova explosions, can be used to make a formal separation between the two. Note that \cite{meaburn80} called large shells around bursts of star formation 'Giant Shells'. These are the 'Superbubbles' of \cite{chu08} and not to be confused with the 'Super-Giant Shells' of \cite{meaburn80}. As the purpose of this article is to explain our detection technique we will not go more deeply into the subject here.

In the present article we have used observations of the Antennae galaxies, where the linear resolution is such that objects smaller than a few tens of parsecs are not readily detectable, so that all our detections in this article are of superbubbles or giant shells, although the method is equally applicable to bubbles or expanding shells, given data on galaxies sufficiently close. Both phenomena have been studied theoretically and observationally for a long time \citep{dyson72,weaver77,heiles79,mccray79,castor75,mccray87}, and are a key aspect of feedback in galaxies. There have been detailed studies analysing their kinematics, mainly in relatively nearby giant HII regions such as 30 Doradus \citep{meaburn84,chu94} or NGC 604 \citep{yang96}, and other regions in the Magellanic clouds and several nearby galaxies \citep{meaburn80,chu86,martin96,naze01} as well as inside the Milky Way \citep{meaburn70,chu82}.

As a superbubble expands and sweeps up the gas it disrupts the dense clouds where the stars are forming, so that star formation in the cluster may terminate with the formation of the superbubble. The effects on the surrounding ISM, however, can trigger further star formation in neighbouring regions \citep{gerola78,mccray87,palous94}, which implies that the superbubbles have a dual effect of quenching further star formation in the cluster that formed them and, at the same time, spreading star formation through the disc as each new star-forming region triggered by the superbubble will create more superbubbles in turn. This means mechanical feedback from HII regions plays a major role in galaxy evolution as it heavily regulates star formation in the disc;  simulations of these effects provide support for this scenario as shown in \cite{dallavec08,ceverino09}.

One important consequence of the cut-off in time of star formation in the cluster as the ISM is swept into a superbubble is that the age of the superbubble coincides closely with the age of the star cluster. This means that if we can measure the dynamical parameters of the superbubble and compare them with an accurate model for its evolution the age of the cluster can be found in a straightforward way, as an alternative to (or complementing) spatially resolved observations of the cluster and fitting models for the stellar radiation such as Starburst99 \citep{leitherer99,sharma11}, or using the Hertzsprung-Russell diagram (which requires resolving individual stars). It also allows studies of cluster ages to be carried out for objects at greater distances where the spatial resolution is limited.

Here we present a method for finding and mapping expanding components of the ionized gas throughout a galaxy or, for very nearby galaxies, within a sub-field. We use a Fabry-Perot spectrometer to map the H$\alpha$ line profile throughout the galaxy and then search for the double peaked signature of expansion symmetrically spaced in velocity with respect to the main peak. It has been found in previous studies \citep{relano05,relano07} that H$\alpha$ profiles of bright HII regions have wings on both sides of the peak emitted by expanding approaching and receding parts of a shell or bubble surrounding the region.

With our instrument we achieve high velocity resolution ($\sim$8 km/s) which allows us to separate the different kinematic components of the profile, and thus to detect the expansion signature on each individual pixel of a two dimensional field.

In the present article we describe the method, giving examples of the results and of derived parameters to showcase its potential.

In section 2 we describe the data to which we have applied the method, giving the relevant technical specifications of the telescope and the instrument and a brief description of the reduction process. 
The method itself and its mathematical foundation is explained in section 3, while in section 4 we explain the results that can be obtained from it and the possible applications. Finally, in section 5 we discuss the advantages and shortcomings of the method and explore possible improvements and applications.

\section{Data description}
Our code is designed to work on data cubes of galaxies with two spatial dimensions and a third spectral or velocity direction. It is optimized for a spectral line profile consisting of a single blended spectral line, to consider each resolved peak a kinematic feature, but there is no inherent problem in using it with broader spectra apart from the requirement to identify the spectral line component corresponding to each contributing peak.

The usefulness of the analysis is heavily dependent on the resolution that can be achieved with the observations, both spatial and spectral, in relation to the typical velocities of the target phenomena. 
In the case of bubbles or superbubbles the velocities are generally in the range 30-150 km/s \citep{chu94,yang96,martin96} for ionized gas and 5-50 km/s for atomic gas \citep{kamphuis91,chakraborti11,griffiths02}, requiring very good velocity resolution in both cases. 
Because of this it is clear that high resolution instruments which take spectra in two dimension, such as Fabry-Perot detectors or aperture synthesis radio telescopes such as VLA and ALMA are ideal in supplying inputs for the kind of multi-peaked profile analysis presented here.

Other methods capable of performing analysis of kinematic components, such as echelle spectrometers can take advantage of the algorithms described here, but those reach their true usefulness with two spatial dimensions, as the number of spatial elements to be analysed grows geometrically. Instruments with one spatial dimension, while also useful for these studies, do not allow a full characterization of the bubble, only along a line usually cutting through the centre of the region.
Stepped longslit or long multislit spectra, without cross-dispersion, and fed by fibre bundles produce similar 3-D line profiles to the Fabry-Perot described here, with the advantage of a much wider spectral range. However the instruments available have fields which, at biggest cover an area an order of magnitude smaller than a Fabry-Perot. For this reason when profiling a single emission line the latter is normally used.

An excellent example of the use of echelles in this context can be found in Meaburn's article on 30 Doradus \citep{meaburn84}

The data used here to develop and apply the method were taken with the integral field spectrometer Galaxy H$\alpha$ Fabry-Perot System (GH$\alpha$FaS, see \cite{hernandez08,fathi08}) mounted at the Nasmyth focus of the 4.2m William Herschel Telescope (WHT) at the Observatorio del Roque de los Muchachos (ORM, La Palma). It produces a 3.4x3.4 arcmin data cube with seeing-limited angular resolution (spaxel size $\sim0.2$ arcsec) and a spectrum at each point on the object, with 48 channels covering a spectral range which depends on the central wavelength of the emission but is usually $\sim400$ km/s, yielding a velocity resolution of $\sim$8 km/s for H$\alpha$.

GH$\alpha$FaS does not use an optical derotator so a correction is applied by software, the process is described in detail in \cite{blasco10}. Velocity calibration, phase correction, sky removal and adaptive binning are performed using the procedure described in \cite{daigle06} to produce the final cube which is used in our study.

An important feature of GH$\alpha$FaS system is the IPCS (Image Photon Counting System) detector, which includes a CCD and a photon intensifier. The main advantage this provides, which under many circumstances more than counter balances the lower quantum efficiency, is that it has zero readout noise, an important property when studying extended features of galaxies with low surface brightness which is especially useful when considering very faint secondary peaks.

\section{Multi-peaked emission analysis}
We aim to detect the presence of multiple components in the emission line profile for each spatial position in the data cube, so we need a way to automatically estimate and fit the components for it to be practical. Otherwise we would have to oversee and interact with a very large number of points (for GH$\alpha$FaS, with a 1024x1024 field it is over a million spaxels), which is impractical. To this end we have developed algorithms included in BUBBLY capable of performing this work automatically which yield reliable and precise results.

\subsection{The second derivative method}
In order to obtain the number of Gaussians for each profile and  good estimates of their parameters, which is necessary to avoid local minima when fitting, we take advantage of the properties of the second derivative of Gaussians, following \cite{Goshtasby}.

\begin{figure}

\includegraphics[width=\linewidth]{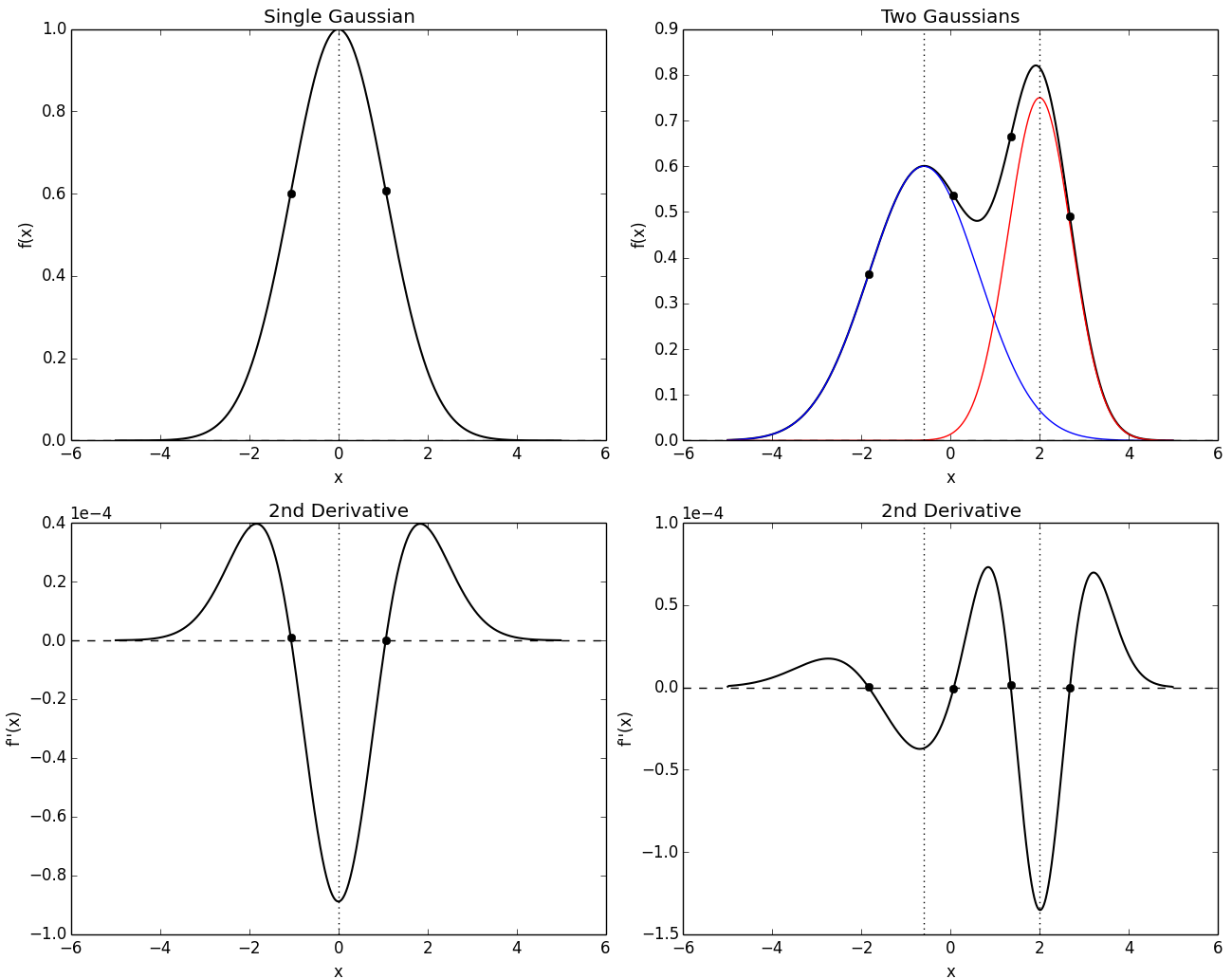}
\caption{Graph illustrating the effectiveness of the second derivative method to separate Gaussian 
components. 
The points correspond to the zeros of the second derivative. Note that the centre of any Gaussian lies between two zeros of the second derivative, a positive to negative zero and a negative to positive zero. 
This holds true regardless of the direction in which you sweep the profile to search for these zeros and is used to find and determine the initial parameters for each Gaussian component.}
\label{fig:gaussexp}
\end{figure}

A single Gaussian, or a normal distribution, $f(x)$, is defined by:

\begin{equation}
 f(x)=\frac{1}{\sqrt{2\thinspace\pi}\sigma}\thinspace e^{-\frac{(x-\mu)^2}{2\sigma^2}}
\label{eq:gauss}
\end{equation}

where $\mu$ and $\sigma$ are the mean and the standard deviation of the distribution respectively.  The second derivative, $f''(x)$, is:

\begin{equation}
 f''(x)=\frac{1}{\sqrt{2\thinspace\pi}\sigma^5}\thinspace e^{-\frac{(x-\mu)^2}{2\sigma^2}}\thinspace 
(\mu^2-\sigma^2+x^2-2\thinspace\mu\thinspace x)
\label{eq:gaussderiv2}
\end{equation}

The second derivative will take the value zero at $x_1=\mu - \sigma$, and $x_2=\mu + \sigma$, so, we can estimate the mean and the standard deviation as:
\begin{equation}
 \mu=\frac{x_1+x_2}{2}
\end{equation}

\begin{equation}
  \sigma=\frac{x_2-x_1}{2}
\end{equation}

This can be generalized to a function comprising N Gaussians:

\begin{equation}
  f(x)=\sum_{i=1}^{i=N}\frac{1}{\sqrt{2\thinspace\pi}\sigma_i}\thinspace e^{-\frac{(x-\mu_i)^2}{2\sigma_i^2}}
  \label{eq:gaussn}
\end{equation}

where $\mu_i$ and $\sigma_i$ are the mean and standard deviation of the i-th Gaussian. 
The second derivative of eq. \ref{eq:gaussn} is:
\begin{equation}
  f''(x)=\sum_{i=1}^{i=N}\frac{1}{\sqrt{2\thinspace\pi}\sigma_i^5}\thinspace e^{-\frac{(x-\mu_i)^2}{2\sigma_i^2}}\thinspace (\mu_i^2-\sigma_i^2+x^2-2\thinspace\mu_i\thinspace x)
\label{eq:gaussderiv2n}
\end{equation}

In order to obtain an initial value of the means and the standard deviations of the Gaussians, and to estimate the ranges of their possible values, we assume that, for any components, $i$, $j$, of the sum of Gaussians of comparable standard deviations, $\sigma_i\sim\sigma_j$, their exponential terms in equation \ref{eq:gaussderiv2n} satisfy:

\begin{equation}
 \frac{1}{\sqrt{2\thinspace\pi}\sigma_j}\thinspace e^{-\frac{(x-\mu_j)^2}{2\sigma_j^2}}>>
\frac{1}{\sqrt{2\thinspace\pi}\sigma_i}\thinspace e^{-\frac{(x-\mu_i)^2}{2\sigma_i^2}}
\end{equation}
for $i\neq j$ and for values of $x$ close to $\mu_j$ and far enough from $\mu_i$. This implies, but only as an initial approximation from which to start the process, that the Gaussians do not overlap, i.e. that they do not contribute significantly to each other's profile. Then:

\begin{equation}
 \mu_i\backsimeq\frac{x_{i,1}+x_{i,2}}{2}
\end{equation}
\begin{equation}
  \sigma_i\backsimeq\frac{x_{i,2}-x_{i,1}}{2}
\end{equation}

where $x_{i,1}$ and $x_{i,2}$ are the pair of zeros associated with the $i$th Gaussian component. 
This approximation clearly works better as the Gaussians are better resolved, but we want to be able to reach the limit of our spectral resolution, so we use them to estimate initial values and possible ranges used in the subsequent fit.

The limits for the mean are the values for the zeros themselves, $x_{i,1} $ and $x_{i,2}$, preventing each component from overlapping completely with the rest:

\begin{equation}
\mu_i^{\mathrm{final}}\in [x_{i,1},x_{i,2}]
\end{equation}

We can also set limits to the standard deviations, but we implement them differently depending on whether the approximate value is greater than one spectral (or velocity) resolution element, $\delta$. 
If it is greater we let it vary between half and double the estimated vale, while if it is lower we use the resolution instead:

\begin{equation}
\sigma_i^{\mathrm{final}} \in \left\{
	\begin{array}{lr}
		\left[\frac{\sigma_i}{2},2 \, \sigma_i \right] & \mathrm{if}\;\sigma_i > \delta \\[1em]

		\left[\frac{\delta}{2},2\thinspace\delta \right] & \mathrm{if}\; \sigma_i < \delta 
	\end{array}
	\right.
\end{equation}

For the intensity we take the initial value as the intensity of the profile at the estimated mean and limit it to between zero intensity and 1.5 times the estimated value.

At this point we have a complete set of detected components of the profile with their respective estimated intensity, offset velocity and standard deviation, which we can feed to the fitting algorithms to obtain the final results.

\subsection{Fitting and results}
We use the properties described above to perform the detection and fitting of the Gaussian components for each spatial point. BUBBLY sweeps through the points which have sufficiently high level of emission, defined by the noise, and performs the following steps: Detecting zeros of the second derivative of the profile, choosing the valid pairs of zeros, estimating the input parameters for the fit and finally performing the fit and saving the results.

A real line profile will have noise which, given its fluctuations, will produce a large number of zeros of the second derivative of the profile. There is also the possibility of an out of range peak extending partially into the range, producing an unpaired zero that would throw off the determination of all the components. To account for this we select the valid pair of zeros by imposing a condition that the in first zero the second derivative goes from positive to negative and vice-versa for the second which is characteristic of a true peak (see Fig. \ref{fig:gaussexp}). We also weed out all pairs between which the profile does not reach an intensity above a threshold signal to noise ratio (SNR).

Once we have the final choice of zero pairs we estimate the initial values and ranges as explained in the previous section and feed them to the algorithm for multiple Gaussian fitting. Our algorithm has been constructed using modified versions of functions specifically developed to fit Gaussians created by Adam Ginsburg for his \textit{gaussfitter} Python program. This program is a wrapper with multiple options (such as upper and lower limits to the parameters) to the \textit{MPFIT} module for Levenberg-Marquardt least squares fitting \footnote{Both \textit{gaussfitter} and \textit{MPFIT} python packages can be found at http://code.google.com/p/agpy/}.

As a result we obtain multi-component line emission surface brightness, velocity and velocity dispersion maps. Since the process is done individually on each spatial point the number of detected components will vary over the field.

These maps can be highly useful for many applications, in this paper we will focus on superbubbles, but there have been a number of different types of studies requiring the analysis of multiple emission peaks. 
A particular topic of interest is the detection and characterisation of  Seyfert galaxies based on the presence of multiple components in the emission line profiles. In \cite{arribas96,garcia13} the double-peaked emission line profiles of NGC 1068 are sought, and used to analyse the active nucleus of the galaxy, while \cite{arribas01} perform a similar study on Arp 220. 
\cite{liu10,ge12,pilyugin12} explore the possibility of using multiple peak analysis on a sample of SDSS galaxies, where automated methods similar to ours would be of great value. 
Another application of the presence of multiple peaks in data cubes is found in \cite{williams94}, where a routine to separate the blended emission of molecular gas clumps is presented. Yet another example is the detection of hydrogen gas transfer between a pair of interacting galaxies in \cite{font11}. 

As an example of the application of our method in particular, in \cite{2013MNRAS.432..998Z} a local inspection of the multi-component velocity maps around the nuclear starburst in NGC 3396 was made. 
The authors detected, using the secondary components of the velocity, an outflow of ionized gas.

\section{Application to expanding structures}
We can easily apply the results from the multi-peak fit to search for expanding structures. The kinematics along the line of sight for an expanding shell around an HII region present three peaks, one in the middle produced by the bulk of the region, one for the gas approaching us because of the expansion, and a corresponding peak receding from us emitted by the far side of the shell. Because of this we define the expansion signature that we will seek in the profiles as a bright main emission peak, coming from the bulk of an HII region, and two secondary peaks placed symmetrically, one at each side of the main peak.

The inhomogeneity of the ISM, the effect of the density profile of the gas decreasing perpendicularly to the disc and instrumental error will generally cause these two velocities to differ slightly from perfect symmetry, so we allow a maximum deviation from symmetry with respect to the central peak of twice the spectral resolution of our instrument, $2\delta v \sim$16km/s. 
Tests were performed comparing the results obtained using different allowed deviations. Initially we used a deviance of $\delta v$, the instrumental error for these measurements, but we found that the $2 \delta v$ maps contained the same features as the $\delta v$ maps, i.e. no additional macroscopic features, and the structures detected for each feature appeared more complete. Further tests using $3 \delta v$ did not yield more coherent structures but did introduce false signatures, so the $2 \delta v$ window was used throughout this work.

Once it has the multiple component maps BUBBLY determines the main peak of the line profile for each spatial point, using flux as the criterion, and then checks the other detected peaks to find a pair that satisfies the conditions imposed. It then creates a map containing the mean separation between these peaks (the expansion velocity at that point) and another map with their mean intensity. For the example shown here we used instead the relative intensity with respect to the total intensity in order to use fluxes obtained from narrow band images ($I_b=0.5*(I_1/I +I_2/I)$).
BUBBLY offers the option of imposing additional conditions on the secondary peaks to accept them as an expansion signature, namely having similar intensities and velocity dispersions for the symmetric pairs.
\begin{figure}
\includegraphics[width=\linewidth]{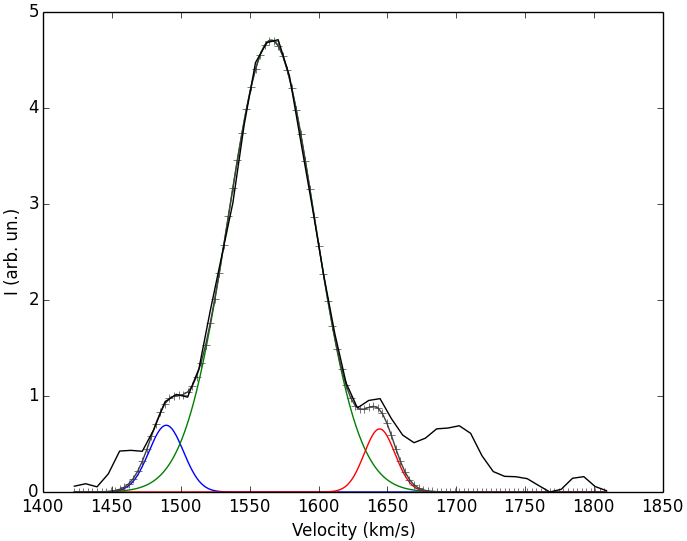}
\caption{Example of expansion detection in an HII region of the Antennae (see Fig \ref{fig:exp_map}). 
The individual components (green, red and blue) are plotted along with the profile (black) and the fit result (crosses). The brightest, main peak corresponds to the emission from the bulk of the HII region, and the two smaller, symmetrically placed peaks on each side correspond to the approaching and receding parts of the superbubble along the line of sight in that particular spatial point. Further peaks can be observed on the profile, but they are not taken into consideration here as they are not symmetric in velocity about the central peak, which they would be in the case of expansion. They represent a complexity of kinematic structure which is clearly present, but which we are, for the present, not pursuing in detail.}
\label{fig:exp_ex}
\end{figure}

Another result of the overall mapping process is a velocity map of the main peak, which gives the overall kinematics of the galaxy. This map is an alternative to the usual method for obtaining a kinematic map in H$\alpha$ which uses the moments method. The latter calculates the moments of the line profile to find the central "characteristic" velocity, while our algorithm takes the velocity of the brightest peak of each profile. In general our algorithm gives a more reliable result, because if there are secondary emission peaks these will weight the moments method and give a distorted value for the velocity of the spaxel concerned. 
However in kinematically complex environments, where there may be a number of peaks with comparable intensities, one can obtain artificial jumps in the velocity map using our algorithm, because it selects the brightest of these peaks, regardless of whether this represents the overall underlying velocity, while the moments method finds an average, and smooths the differences.
In this section we have included an example of the application of BUBBLY to a data cube and the results that originate from it to demonstrate the detection capabilities of the program. The selected object is the Antennae pair of galaxies, which exhibits very interesting results on superbubbles originating from powerful current star formation. A paper analysing the superbubbles in depth and comparing the detections to previous work on them is in preparation.

\subsection{Finding the expanding structures}
\begin{figure*}
\includegraphics[width=\linewidth]{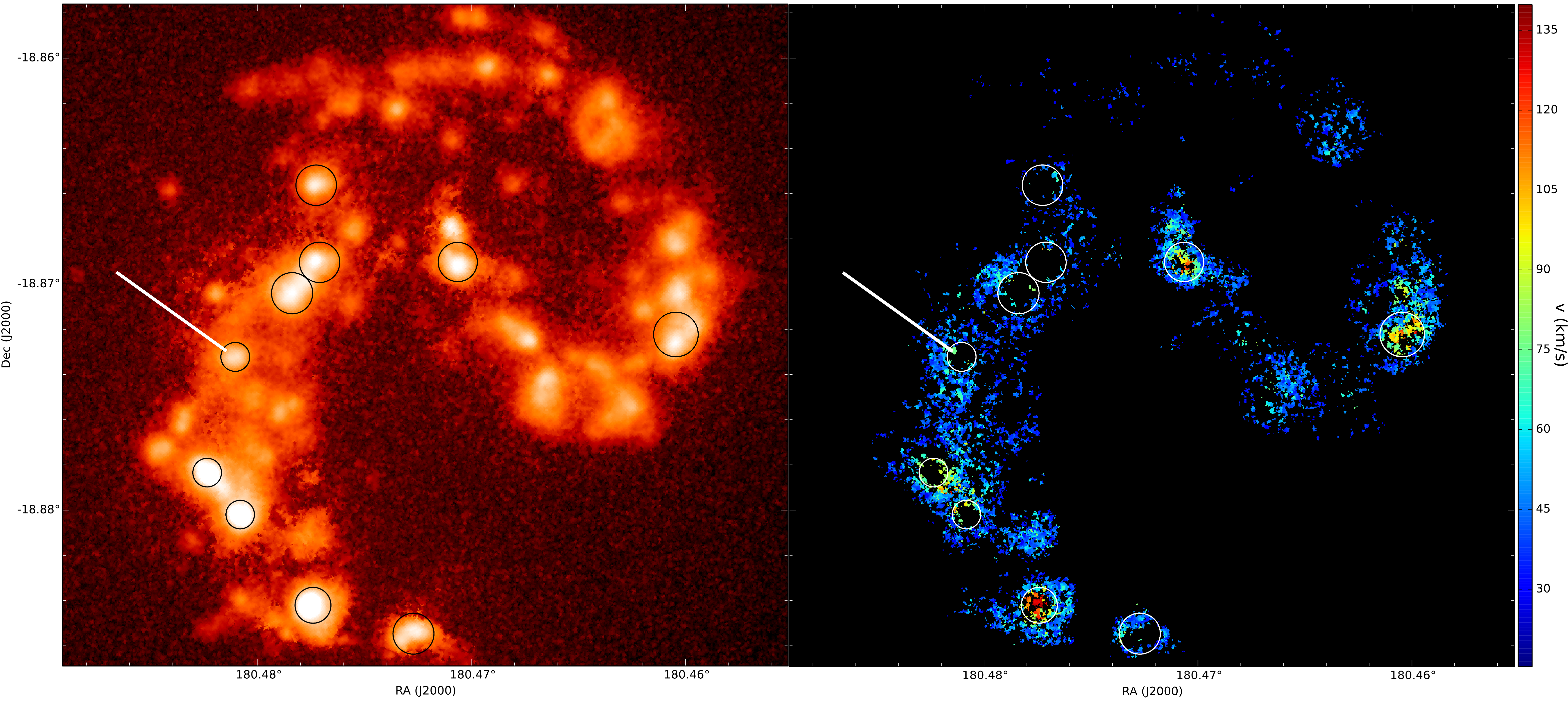}
\caption{Comparison between the H$\alpha$ narrow band image (left) of the Antennae galaxies and the corresponding expansion map (right). The circles represent the estimated bubble radii and the white line signals the position for the H$\alpha$ profile shown in Fig. \ref{fig:exp_ex}. The scale is in velocity with respect to the emission from the bulk of the HII region.}
\label{fig:exp_map}
\end{figure*}

After processing the data cube with BUBBLY the main products we obtain are the expansion velocity map, the intensity map and velocity, the intensity and dispersion maps for the main peak. The first two are the relevant ones for the study of superbubbles, as they are what we will use to detect and characterise them.
In Fig. \ref{fig:exp_map} we show the expansion map of the Antennae galaxies and its corresponding H$\alpha$ narrow band image as an example of the products of our method. It can be seen in the image that the centres of the superbubbles detected coincide well with the centres of the brightest HII regions. 
A paper analysing the properties of the superbubbles in this object will be submitted in the near future.

\begin{figure}
\includegraphics[width=\linewidth]{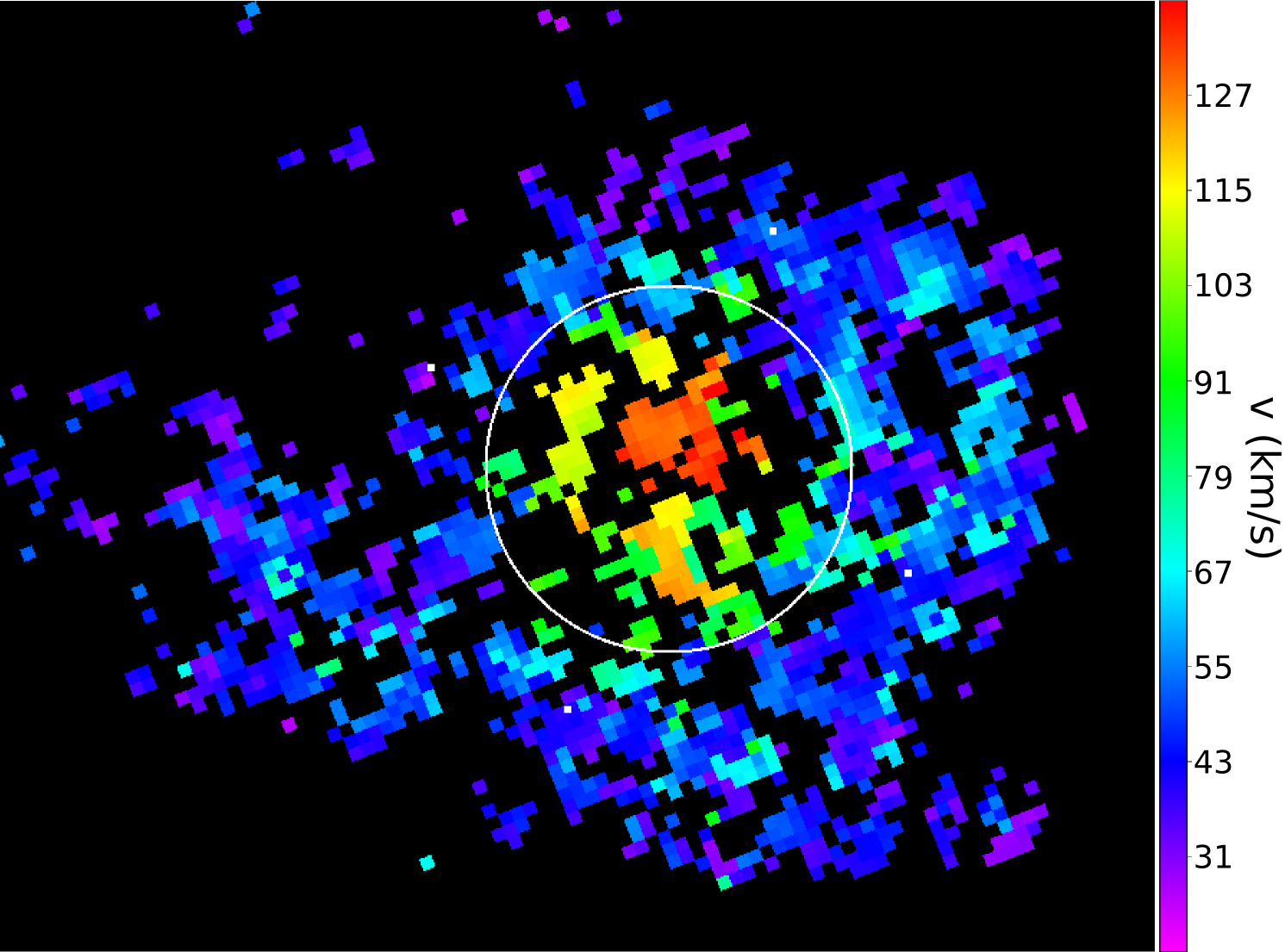}
\caption{Detail of a superbubble detected in the brightest HII region in the Antennae galaxies. The scale is velocity with respect to the emission from the bulk of the HII region and the white circle represents the estimated superbubble radius. As expected from an expanding structure, we can see the highest velocity at the centre, falling to lower values towards the edge. The spatial coherence and effect of projection on the observed velocity can be well seen in this example.}
\label{fig:exp_bub}
\end{figure}
We detect the superbubbles by inspecting the expansion map to find regions with coherent patterns in the expansion velocity that correspond to the expected radial distribution of observed expansion velocity characteristic of superbubbles. These arise from the fact that the velocities we measure are the components directed along the line of sight due to the Doppler effect, so a spherically expanding shell will show projection effects on the expansion map. Therefore, observed expansion velocity will be highest at the centre of the bubble where the velocity vector aligns with the line of sight, decreasing radially until it is zero at the edge, so we look for radially decreasing circular patterns in the map.

A particularly clear case of a detected superbubble from the Antennae is shown in Fig. \ref{fig:exp_bub} with the estimated bubble size and position shown by a white circle. This example is the clearest one on the object, because usually we do not detect expansion in the very centre of the shell (which is typically optically thin) but only an annulus concentric with the HII region.

The absence of observable expansion components in the central zone of a superbubble is due to a combination of phenomena. One is that superbubbles of this size enter a blowout phase as they grow beyond the vertical density scale height of the disc, so that the gas of the central part thins out as it expands freely into the halo; another is the extinction caused by dust which may obscure the receding component of the bubble signature, as that emission has to travel through the entire region to get to us, and the centre of the far face of the superbubble is the part with the largest optical path to the observer. 

Both of these factors are combined with the aforementioned fact that the observed column density for a spherical shell is lowest at the centre because of the projection effect.

Naturally, the assumption of sphericity is inexact at best. There have been many theoretical and simulation studies of the evolution of such superbubbles \citep{maclow88,zaninetti04}, all of them concluding that they soon acquire an elongated shape, with the long axis  perpendicular to the plane of the galaxy, as the ambient density decreases in that direction, morphing into a peanut shape for bigger and older superbubbles. The projection effects described still apply, but it can be expected, for example, that the observed projected expansion velocity will be affected. As the superbubble has a peanut shape its observed edge as seen from above will not have a zero value of the vertical velocity, so we still see the gradient, but the expansion velocity will decrease more slowly and we will stop detecting the bubble before it approaches zero. The expansion velocity profiles we have found do agree better with the elongated case.

\subsubsection{Multiple expanding features}
\begin{figure}
\includegraphics[width=\linewidth]{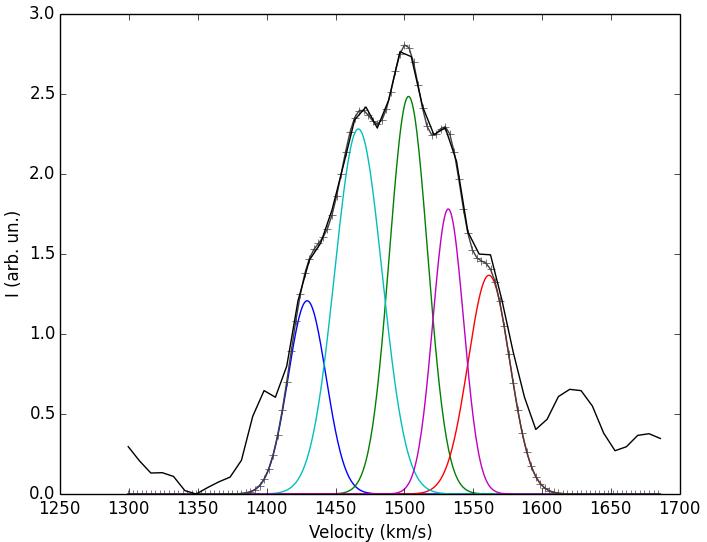}
\caption{Profile observed in the Antennae galaxies where two pairs of symmetrically spaced secondary peaks are detected, implying the possible presence of two superbubbles in one region. An alternative possibility is that it could be a projection effect with two HII regions overlapping on the line of sight which may indicate more complex kinematics.
As in Fig. \ref{fig:exp_ex} we show separated peaks only for detected components that are symmetrical to the main peak.}
\label{fig:double_bub}
\end{figure}

In some cases around the brightest HII regions, profiles exhibiting multiple pairs of symmetrically separated peaks can be observed (Fig. \ref{fig:double_bub}). These additional features follow the same general shape as the main detected superbubbles, forming rings around the HII regions, but they are not as common and the analysis and determination of their properties is more difficult.

However, they represent a very interesting possibility as their presence may imply multiple bursts of star formation in the same regions or perhaps a system of more local bubbles inside the large region and its associated superbubble.
Another possibility to explain the presence of multiple expansion signatures in a spaxel would be that we are detecting kinematics which are considerably more complex than that of an assumed superbubble.

Deeper spectral imaging should help to characterize these and other complex features.

\subsection{Bubble properties}
The amount of information about a bubble or shell that we can extract is critical for our method to be truly useful. We have four primary quantities from which we may derive the physical parameters of a particular bubble: the radius, the HII region luminosity, the shell to region luminosity ratio, and the expansion velocity of the shell.
We can calculate the bubble luminosity, density, mass, kinetic energy and age.
The basic equation we use is the luminosity of a shell \citep{relano05}

\begin{equation}
L_{H{\alpha}}(shell)=4\pi R^2 \Delta R n_{e}^2 \alpha(H,T) h \nu
\label{lum_eq}
\end{equation}

Among the parameters appearing in this equation are the effective recombination coefficient for hydrogen $\alpha^{\mathrm{eff}}_{H\alpha }(H,T)$, which can be taken as $4 \cdot 10^{-13}$  for HII regions \citep{osterbrock89}, $h\nu_{H\alpha}$ is the energy of an H$\alpha$ photon and the shell width $\Delta R$ which we take as $\sim$2pc. We do not expect the width to vary much from case to case and when calculating the mass and subsequent quantities its contribution varies as $\Delta R ^{-1/2}$, so it does not affect the final result very strongly. Observational uncertainties give a much larger possible variation so we can safely consider it as constant.

Since we already know the luminosity of the shell from the product of the region luminosity and the shell-to-region luminosity ratio we are especially interested in the density to calculate the total mass of the shell and its energy. We can also make a zero order estimate of the age of the bubble by dividing the radius by the expansion velocity, which is equivalent to assuming the region has expanded at constant velocity. Thus:

\begin{equation}
n_e=\sqrt{\frac{L_{H{\alpha}}(shell)}{4\pi R^2 \Delta R \: \alpha^{\mathrm{eff}}_{H\alpha }(H,T) \nu_{H\alpha}}}
\label{dens_eq}
\end{equation}

\begin{equation}
M=4\pi R^2 \Delta R \: n_e m_p
\label{mass_eq}
\end{equation}

\begin{equation}
E_k=\frac{1}{2}M*v_{exp}^2
\label{energy_eq}
\end{equation}

\begin{equation}
Age= R/v_{exp}
\label{eq_age}
\end{equation}

In Table \ref{prop_table} there is a list of these measured parameters for some of the superbubbles we have 
detected in the Antennae. The errors for the parameters are determined by propagation of the errors in 
radius, expansion velocity and H$\alpha$ luminosity. The first two parameters are estimated by us directly 
by inspection of the expansion map, so we took a conservative error of 10\% for both radius and expansion 
velocity. For the luminosity the most important source of error is, by far, the uncertainty in the radius 
of the region, though it does not propagate in a mathematically straightforward way as it affects the pixels 
over which we sum the detected flux and depends on the physical characteristics of each region. What we do 
is measure three sets of fluxes for each region, two of which correspond to the high and low extreme values 
of the radius as allowed by the radius error, and take their mean deviation from the central value as the 
error for the flux.

\begin{table*}
\centering
\caption{Properties of some detected bubbles in the Antennae galaxies}
\begin{tabular}{ccccccc}
\hline \hline
Radius & $v_{exp}$ & L$_{\mathrm{H}{\alpha}}$(shell) & $n_e$ & Mass & Kinetic energy & Age \\
(pc) & (km/s) & $10^{38}$ erg/s & (cm$^{-3}$) & ($10^{35}$ Kg) & ($10^{51}$ erg) & Myr\\
 (1) & (2) & (3) & (4) & (5) & (6) & (7) \\
\hline
334	$\pm$	33	&	115	$\pm$	12	&	3.04	$\pm$	0.25	&	6.4	$\pm$	0.7	&	7.36	$\pm$	1.5	&	48.7	$\pm$	13.6	&	2.8	$\pm$	0.4	\\
304	$\pm$	30	&	82	$\pm$	8	&	0.44	$\pm$	0.06	&	2.7	$\pm$	0.3	&	2.55	$\pm$	0.5	&	8.6	$\pm$	2.6	&	3.6	$\pm$	0.5	\\
319	$\pm$	32	&	73	$\pm$	7	&	1.83	$\pm$	0.21	&	5.2	$\pm$	0.6	&	5.45	$\pm$	1.2	&	14.5	$\pm$	4.5	&	4.3	$\pm$	0.6	\\
415	$\pm$	42	&	82	$\pm$	8	&	0.77	$\pm$	0.10	&	2.6	$\pm$	0.3	&	4.59	$\pm$	0.9	&	15.4	$\pm$	4.4	&	4.9	$\pm$	0.7	\\
269	$\pm$	27	&	70	$\pm$	7	&	0.16	$\pm$	0.02	&	1.8	$\pm$	0.2	&	1.36	$\pm$	0.3	&	3.3	$\pm$	0.9	&	3.8	$\pm$	0.5	\\
383	$\pm$	38	&	57	$\pm$	6	&	0.80	$\pm$	0.10	&	2.9	$\pm$	0.3	&	4.33	$\pm$	0.9	&	7.0	$\pm$	2.0	&	6.6	$\pm$	0.9	\\

\hline
\end{tabular}
\\
\begin{flushleft}
\textbf{Notes.} All parameters refer solely to the bubble itself, not the underlying HII region. The 
kinetic 
energy (7) is given in units of $10^{51}$ erg, which is approximately the energy one supernova injects 
into the medium. In this manner, the energy can be taken to be in units of supernovae necessary to drive 
the expansion in that particular region. The age (8) is calculated as R(2)/$v_{exp}$(3).

\end{flushleft}
\label{prop_table}
\end{table*}

\subsubsection{Age dating}
A precise determination of the age of a superbubble is very important as it is equivalent to the age of the 
associated HII region and can therefore be used to give a short-term map of the star forming history of the 
galaxy. In Table \ref{prop_table} we list ages for each superbubble, but these are merely intended as rough 
approximations given the crude way in which they are calculated, as simply $R/v_{exp}$. 
The superbubble evolves as a complex interaction of the winds, supernovae and ionization pressure with the 
non-isotropic interstellar medium, and the expansion velocity will vary throughout its lifetime.

A better approach to the issue is therefore necessary and entails models of the interaction. 
We have experimented with analytic models for the expansion of superbubbles with moderate success, comparison 
with observations of line equivalent widths in clusters to produce their stellar ages gives compatible ages 
in most cases, especially for the younger regions. The models assume homogeneous distribution of gas in 
all directions, while in reality and on the scales we are studying, it becomes necessary to take into 
account the 3D shape of the distribution. We are working with detailed physical simulations to provide 
a better determination of the ages and other parameters.

\section{Discussion}
We have presented our method and associated software designed to detect and fit multiple components to 
integral field spectroscopy of emission lines, with an application to superbubbles using our Fabry-Perot 
instrument.

In section 3 we have explained in detail the method for the line profile decomposition into a set of 
Gaussians and the way it is implemented in BUBBLY. The next section applied the results from the 
decomposition to produce expansion maps that can be used to search for and characterise superbubbles. 
In this regard with a relatively simple approach we already obtain reliable detection and characterization 
of superbubbles.

The presence of multiple components in an emission line is a powerful way to study the fainter, and more 
difficult to detect, kinematic phenomena in a galaxy, as well as numerous other applications such as 
separating the emission of different regions or clouds or even galaxies that overlap or AGN and violent 
galactic outflows (see \S 3.2). Our algorithm has the advantage of being both simple and easy to use and 
very flexible with regards to the input data, which makes it a potent tool to study galaxy kinematics.

In this paper we have centred on the application of multi-component analysis to the search and 
characterisation of superbubbles, where the method proves especially useful allowing us to detect several 
superbubbles clearly associated with giant HII regions. The fact that it does not need purpose-programmed 
observations to work means it offers added value to integral field spectroscopy observations of galaxies.

Current alternative methods for the detection of superbubbles are to use morphological features observed 
through narrow H$\alpha$ filters such as cavities and arches around OB associations, shock-fuelled 
forbidden emission lines such as SII and OIII and morphological and kinematic features in HI.
 X-ray observations of extended emission interior to the observed shells is a complementary diagnostic.

All morphology-based methods require very high spatial resolution and  shells which are bright with respect 
to their surroundings in order to detect them, which restricts the methods to nearby galaxies and/or 
observations with very high angular resolution. This is not the case for our method, where the shell 
emission is separated from the HII region emission kinematically, although of course with better angular 
resolution we will be able to detect more and smaller bubbles.

Except for HI cube observations the other methods do not provide kinematics of the shell which prevents 
a detailed characterization of the bubble and its dynamic properties, limiting the utility of the data. It 
would be possible to complement the observations with high-resolution spectra to detect the expanding 
components but the total observing time required would be very high, considering that each region would 
require an individual spectrum. 

The most direct competition for our method when it is applied to superbubbles is, therefore, observing HI 
using the classical 21 cm method. The long wavelength of this HI emission makes it very difficult to achieve 
the spatial resolution necessary to distinguish these features properly, (even though the velocity resolution
is high), making it very difficult to extend these studies beyond the closest galaxies in the local group, 
such as M101 where several shells have been detected in this manner \citep{kamphuis91,chakraborti11}. 

In any case the shells detected with HI normally correspond to the late stages of the superbubble, after the 
massive stars which ionized the shell have disappeared and the superbubble has decelerated greatly. Because of 
this the typical expansion velocities lie in the range 5-50 km/s \citep{kamphuis91,chakraborti11,griffiths02}
 and in many cases it is difficult or impossible to find the OB association corresponding to the superbubble. This
 is not the case for our method which traces the superbubbles from the ionized gas, so we find young superbubbles 
clearly associated with the cluster. In this manner, it can be said that HI observations and ours are 
complementary, each detecting superbubbles in different stages of their evolution.

Of course, the method for multi-peak detection can be applied to HI data cubes without problems, but to 
detect superbubbles we would have the problem of lacking a main peak associated with the emission of the 
region and therefore a reference velocity for the secondary peaks. This could be solved using the velocity 
field of the galaxy for comparison, which is in fact a normal product of the analysis of this data.

 It would certainly be possible to adapt our code to be run on HI data cubes to apply the method and yield 
the expansion maps.

We are limited in the range of application by the faint nature of the secondary components we try to detect 
and the spatial resolution necessary to see the structure of the superbubble, which is required in order to 
decouple the shell signature from complex non-expanding kinematic structure along the line of sight.

The first problem prevents expansion detection at some points where one of the components is not observed, 
though careful examination of the profiles of the regions clearly shows that in most of these points one of 
the pair of secondary peaks that make up the expansion signature appears, or even both, with one of them 
having a lower intensity and falling below the minimum SNR required by the program. These appear at the same 
velocities as in the points with double detection, so it is evident that they correspond to the same expanding structure.

The second problem (spatial resolution) impedes a more complete and statistical study of all detected 
expansions, as we need to see the characteristic projection structure of the velocity. Otherwise we could 
distinguish real signatures and their properties at much smaller scales, yielding many more bubbles/shells per 
galaxy and allowing a much more detailed study of the feedback effects within the galaxy.

The values for the parameters we obtain (see Table \ref{prop_table} for some examples) appear to be 
reasonable and consistent among the regions. The results should be used with some care, however, as they 
rely on several simplifications such as assuming a perfect spherical shell and a homogeneous medium, which 
we know to be only moderate approximations. We expect these deviations to be important but not critical, 
so the properties we have obtained should be an acceptable approximation for many purposes.

By far the biggest source of uncertainty when deriving these parameters is the radius of the shell as it 
influences strongly the determination of the flux and it also appears as an important parameter for the 
calculations of the physical parameters. This is because the determination is performed by inspection and 
suffers from a lack of spatial coherence in many cases, so we chose a conservative margin of error to be safe. The value of the radius is also subject to some uncertainty because the true shape of the superbubble is 
non-spherical. We expect superbubbles to be elongated in the direction perpendicular to the disc, and 
therefore we are usually measuring the lesser radius of an ellipsoid. This would mean we are underestimating 
the kinetic energy by a factor depending on how elongated the structure is, but another effect acts in the 
opposing sense. In an elongated superbubble the expansion along the major axis (perpendicular to the disc) is 
faster than along the minor axis (parallel to the disc), this means that the expansion velocity we use is an 
upper limit and thus we are overestimating the kinetic energy. The combination of the two effects should 
tend to cancel somewhat, but in general younger and smaller superbubbles will be more spherical and thus 
provide more reliable energy and age determinations.

We are working to use hydrodynamic simulations to constrain these effects and obtain more reliable parameter 
determinations.

\subsection{Further developments and applications}
Throughout the text the current shortcomings of the method and possible developments to address them and 
further improve the results are commented. Here is a summary of the main possible developments and 
applications:

\begin{enumerate}

\item Write additional modules to take advantage of the multiple component emission detected in order to 
search for wind streams, inflows and outflows within a galaxy.

\item Improve the detection capabilities of the code by implementing pattern recognition algorithms in order 
to correlate the signatures between themselves and detect lower emission signatures, allowing us to better 
characterise bubbles/shells and to expand the study to smaller scales.

\item Obtain a precise determination of the age of the expanding structure and, subsequently, the underlying cluster 
by means of hydrodynamic simulations.

\item Explore the possibility of the presence of multiple expansions within a region and the implications with 
regards to the conditions determining  star formation.

\item Apply results obtained from our method to study feedback and evolution in galaxies.

\end{enumerate}

The feasibility of this program is based on the effectiveness of BUBBLY as a superbubble detector.

\section*{Acknowledgements}
This research has been supported by the Spanish Ministry of Economy and Competitiveness (MINECO) under the grant AYA2007-67625-CO2-01, and by the Instituto de Astrof\'{i}sica de Canarias under project P/308603. JEB acknowledges financial support to the DAGAL network from the People Programme (Marie Curie Actions) of the European Union's Seventh Framework Programme FP7/2007-2013/ under REA grant agreement number PITN-GA-2011-289313.

The William Herschel Telescope is operated on the island of La Palma by the Isaac Newton Group in the Spanish Observatorio del Roque de los Muchachos of the Instituto de Astrof\'{i}sica de Canarias. 

We thank the referee, John Meaburn, for very helpful comments which have enabled us to make
a significant improvement in the paper.

     \bibliographystyle{mn2e}
    % \bibliography{references/all_references.bib} 

\label{lastpage}

\end{document}